\documentclass[conference]{IEEEtran}
\IEEEoverridecommandlockouts

\usepackage{cite}
\usepackage{amsmath,amssymb,amsfonts}
\usepackage{algpseudocode}
\usepackage{multirow}
\usepackage{algorithmicx}
\usepackage{algorithm} 
\usepackage{graphicx}
\usepackage{textcomp}
\usepackage{xcolor}
\usepackage{url}
\usepackage{bm}
\usepackage{subfig}
\newcommand\blfootnote[1]{%
  \begingroup
  \renewcommand\thefootnote{}\footnote{#1}%
  \addtocounter{footnote}{-1}%
  \endgroup
}
\newcommand{\rev}[1]{#1}
\newcommand{\revi}[1]{#1}

\def\BibTeX{{\rm B\kern-.05em{\sc i\kern-.025em b}\kern-.08em
    T\kern-.1667em\lower.7ex\hbox{E}\kern-.125emX}}
\begin{document}

\title{Online waveform selection for cognitive radar
\thanks{This work was funded by DRDO, Ministry of Defense.}
}

\author{
\IEEEauthorblockN{Thulasi Tholeti \IEEEauthorrefmark{2} \IEEEauthorrefmark{1} }
\IEEEauthorblockA{\textit{Institute for Experiential AI}\\
\textit{Northeastern University}\\
Boston, USA \\
t.tholeti@northeastern.edu}
\and
\IEEEauthorblockN{Avinash Rangarajan \IEEEauthorrefmark{2}}
\IEEEauthorblockA{\textit{DYSL-CT Lab} \\
\textit{DRDO}\\
Chennai, India \\
avinash.r@dysl-ct.drdo.in}
\and
\IEEEauthorblockN{Sheetal Kalyani}
\IEEEauthorblockA{ \textit{Dept. of Electrical Engineering} \\
\textit{Indian Institute of Technology Madras}\\
Chennai, India \\
skalyani@ee.iitm.ac.in}

}

\maketitle
\blfootnote{\IEEEauthorrefmark{2} Authors contributed equally. \IEEEauthorrefmark{1} Work done at IIT Madras.}
\begin{abstract}
\rev{Designing a cognitive radar system capable of adapting its parameters is challenging, particularly when tasked with tracking a ballistic missile throughout its entire flight. In this work, we focus on proposing adaptive algorithms that select waveform parameters in an online fashion. Our novelty lies in formulating the learning problem using domain knowledge derived from the characteristics of ballistic trajectories. We propose three reinforcement learning algorithms: bandwidth scaling, Q-learning, and Q-learning lookahead. These algorithms dynamically choose the bandwidth for each transmission based on received feedback. Through experiments on synthetically generated ballistic trajectories, we demonstrate that our proposed algorithms achieve the dual objectives of minimizing range error and maintaining continuous tracking without losing the target.}
\end{abstract}


\section{Introduction}
Cognitive radars are capable of adjusting their parameters in real-time to optimize performance, making them increasingly valuable for military, civilian, and scientific applications \cite{haykin2006cognitive, haykin2010cognitive,bell2015cognitive}. In cognitive radar systems, majority of the error in estimation can be reduced by improving two components: (i) Tracking: \rev{Predicting the position of the target based on the measurement of range, range rate, Azimuth angle and elevation angle and minimizing prediction error. }Popular filtering techniques such as the Gaussian particle filtering \cite{yu2017multiple, liu2011interacting} or Kalman filtering variants can be used to perform tracking. For this work, we assume that an Extended Kalman Filter is used. \cite{kumar2017adaptive}.
(ii) Waveform selection: Choosing the optimal waveform parameters for illuminating the target \rev{at the predicted position}.

     A critical challenge in these systems is the adaptive selection of waveform parameters in response to varying target and environmental conditions. \rev{Choosing waveform parameters that cause a wide window of illumination can result in high range error, whereas a narrow window may miss the target and result in the loss of track.} Traditional approaches rely on predefined strategies or heuristic rules, which may not be sufficiently robust in highly dynamic and uncertain scenarios.
     
     The work in \cite{haykin2010cognitive} propose the use of dynamic programming to adaptively choose the parameters of the waveform. It was proposed that the state space consist of 9 elements (position, velocity, acceleration for each of the parameters: range, azimuth and elevation). The state space is of dimension $n^9$, which is difficult to train even for $n=5$. The adaptive waveform selection was further extended for tracking multiple targets in the work by \cite{zhang2018adaptive} and \cite{xin2022adaptable}. Another approach that uses reinforcement learning for adaptive waveform selection was introduced in \cite{zhu2023cognitive} where entropy reward Q-learning was proposed. 
     
     \rev{\revi{While these other works focus on a general cognitive radar problem,} this work focuses on the tracking problem of ballistic missile in different phases during its flight \cite{benavoli2007tracking}. The trajectory of missiles includes boost, mid-course, and terminal phases. Typically, the maximum estimation uncertainty occurs during the boost phase due to factors like engine performance and atmospheric conditions, while the mid-course phase has relatively lower uncertainty and the terminal phase again see increased uncertainty due to atmospheric effects and maneuvering capabilities of the missile. This becomes a challenge for the radar to optimize the waveform selection such that track accuracy is maintained without loss of track in all three phases of missile. }\rev{Here, we attempt to optimize the waveform parameters, especially the bandwidth, accounting for varied levels of uncertainty from measurement and the tracking in such a manner that the radar can utilize the advantage of higher accuracy, and at the same time ensure track is maintained throughout the trajectory. To this end, we propose using model-free reinforcement learning methods that is tailored for this specific challenging problem.}
\section{System Model}
	The system model is adapted from \cite{haykin2010cognitive}. The transmitted signal from radar is of the form,
	\begin{align} \label{eqn:tx}
	    s_T(t) = \sqrt{2} \text{Re}\left( \sqrt{E_T} \tilde{s}(t) \exp(i 2 \pi f_c t) \right),
	\end{align}
	where $f_c$ is the carrier frequency, $T$ is the time period, $E_T$ is the transmitted energy of the signal and $\tilde{s}(t)$ is the complex envelope. For our system, we consider Linear Frequency Modulation (LFM) where the transmitted signal is, 
	\begin{align}
	    \tilde{s}(t) &= \exp(\left( -j \pi \dfrac{b}{\tau} (t^2- 2 \tau t)\right) 
	\end{align}
	with $|t| \leq T/2 + t_f$ where $t_f << T/2$ is the rise and fall time, $b$ denotes the bandwidth and $\tau$ is the pulse duration. The Pulse Repetition Frequency (PRF) $\lambda$ is inversely proportional to the pulse duration $\tau$. Let $\bm{\theta} = [\lambda, b]^T$ denote the wave parameter vector that needs to be optimized by the adaptive algorithm.
	
	Now, we describe the discrete-time state space model of the system. The state equation of the system is
	\begin{align}
	    \bm{x}_k = \bm{f}(\bm{x}_{k-1}) + \bm{v}_k,
	\end{align}
	where $\bm{x}_k$ is the state of the target at time $k$ and $\bm{v}_k$ is an i.i.d process with mean $\bm{0}$ and covariance $\bm{Q}_k$. The measurement equation for the system is
	\begin{align}
	    \bm{z}_k = \bm{h}(\bm{x}_k) + \bm{w}_k(\bm{\theta_k}),
	\end{align}
	where waveform dependent measurement noise $\bm{w}_k(\bm{\theta_k})$ is modeled as i.i.d Gaussian process with $\bm{0}$-mean and covariance matrix $R_k(\bm{\theta}_k)$.
		At any time $k$, the information available to the transmitter for decision making can be denoted by the information vector $\bm{I}_k = \left(\bm{z}^{k-1}, \bm{\theta}^{k-1} \right)$,
	where $(\bm{z}^{k-1},\bm{\theta}^{k-1}) = \left(\bm{z}_0, \bm{z}_1,\ldots, \bm{\theta}_0, \bm{\theta}_1, \ldots \right)$. 
 The objective of the adaptive waveform selection algorithm is to select the next set of waveform parameters $\bm{\theta}_k$.

\section{Trend study}
To understand the impact of the choice of bandwidth parameters on the performance of the tracking system, we conduct an initial trend study where we vary the parameters of interest and observe the performance. Our metrics of interest during the observation are error in range and the continuity of maintaining track without losing it. Ideally, the target should be continuously tracked throughout its trajectory, ensuring it remains in focus at all times without any loss of tracking. \rev{A track is considered to be lost if there is no correlation between the prediction and the measurement vector of the range parameter for 5 consecutive transmissions.} \rev{(This is a configurable parameter, we choose to declare that the target is lost if there are 5 continuous uncorrelated transmissions.)} Our trend study resulted in the following observations.

The error remains similar when different PRF values are used in transmission. We also note that the use of consistently low PRF values lead to frequent loss of track. The bandwidth has a direct influence on both the range error as well as the tracking. As bandwidth increases, the range error as well as its variance decrease significantly. However, using a higher bandwidth leads to higher chances of the target being lost due to higher uncertainty in target profile at an earlier stage in the trajectory, which is not acceptable.

Due to these observations, we focus on bandwidth as the primary parameter of interest. \rev{This trend study is especially consequential as it helps us observe the performance for our trajectory of interest, which does not exhibit uniform behaviour through its different phases (boost, mid-course and terminal). The work is challenging as the proposed adaptive method for waveform selection should learn to perform optimally in all the three phases, which have varying levels of estimation uncertainty and risk of the track being lost. } 

\section{Proposed methods}\label{sec:prop}
In this section, we propose different methods that adaptively choose the values of the bandwidth $b$. Due to the dynamic nature of the optimization problem, we formulate the problem in the online learning setting, where the algorithm chooses $\bm{\theta}$ for each transmission, receives performance feedback and adapts before choosing the next set of parameters. 
The proposed methods are explained in detail below.

\subsection{Bandwidth scaling method}
 In this method (given as Algorithm \ref{alg:bw_scaling}), we monitor the correlation between the filter tracking and the measurement. We suggest using a high bandwidth for as long as the correlation exists. If the correlation is lost, i.e., \rev{there is zero correlation between the predicted and the measured range vector, the next waveform is transmitted at a lower bandwidth.} Note that the track is determined to be lost only if no correlation exists for over 5 transmissions, \rev{while we keep switching to lower bandwidths after every one of those transmissions with no correlation at avoid loss of track}. 

\begin{algorithm}  
    \caption{Bandwidth scaling algorithm}    \label{alg:bw_scaling}
    \begin{algorithmic}[1]
        \State \textbf{Input}: MAX-BW, MIN-BW, No. of transmissions T
        \State \textbf{Initialization:} $bw_0 =$ MAX-BW
        \For{ \(i = 1,2,\cdots T\)} 
            \If{Correlation = 0}
                \State $bw_{i}=\dfrac{bw_{i-1}}{2}$
                \State Ensure $bw_i \geq$ MIN-BW
            \EndIf
            \If{Correlation = 1 for last 5 transmissions}
                \State $bw_{i}=2bw_{i-1}$
                \State Ensure $bw_i \leq$ MAX-BW
            \EndIf
        \EndFor
        \State Transmit with bandwidth $bw_i$.
    \end{algorithmic}
\end{algorithm}

\subsection{Q-Learning}
In bandwidth scaling, we simply scaled the bandwidth according to the observed tracking performance. This can be formalized by using a learning algorithm which associates each possible action with a value which is learnt over multiple runs. Hence, we propose to employ Q-learning to determine the bandwidth. We formulate the 3 quantities of the learning problem: states, actions and rewards. \rev{Our novelty lies in the formulation of these quantities that is tailored to the ballistic missile trajectory, that yield satisfactory results through the entire course of the transmission. We also prioritize formulating a state space such that the computational overhead is minimal, as this algorithm is employed in real-time before every transmission waveform is sent.}

To formulate our states based on the uncertainty in the prediction and measurement, we use the variances of the prediction and measurement error matrices. \rev{The choice of using the variances of the prediction and measurement error matrices is rooted in our motivation to develop an algorithm that adapts to varying levels of uncertainties. They are further grouped together in intervals; the intervals are designed to minimize computational overload.} If the prediction error variance is grouped into $x$ intervals and the measurement error variance into $y$ intervals, the formulation has $xy$ states. The various possible actions taken by the learning agent are represented by the action set $\mathcal{A}$ whose cardinality is $|\mathcal{A}|$. The reward is calculated using the error in range and if the trajectory is lost. 
\begin{equation}\label{eqn:rewardql}
    r_t = \begin{cases}
            - RangeError & \text{Trajectory not lost}\\
            - C & \text{Trajectory lost}
            \end{cases}
\end{equation}
The constant $C$ can be adjusted based on the desired penalty for losing track, and is typically high in magnitude to enforce a large penalty. To implement the Q-Learning algorithm, a Q-table of dimension $xy \times |\mathcal{A}|$ is constructed and initialized with zeros. The values in the Q-table are called Q-values. Initially, one of the actions is chosen at random and a transmission (step) is made. The Q-table at step $t$ is updated as
\begin{align}
    Q[s_{t-1},a_{t-1}]&=Q[s_{t-1},a_{t-1}]+ \nonumber\\
    &\alpha(r_t + \gamma \max(Q[s_t,:]) - Q[s_{t-1},a_{t-1}]) 
\end{align}
Here, $s_t$ and $a_t$ refer to the state and action at step $t$, $\alpha$ and $\gamma$ are hyperparameters referring to the learning rate and discount factor. Once the Q-table is updated, the action for the next transmission is to be determined. We implement two different action selection methods. (i) Epsilon-greedy method: This action selection method is employed during training. This ensures that there is a trade-off between exploitation (choosing the action known to perform best so far) and exploration (trying other actions). The action with the highest Q-value for a given state is chosen with probability $1-\epsilon$ and a random action is sampled with probability $\epsilon$. Here, $\epsilon$ is known as the exploration factor and controls the degree of exploration. (ii) Best action method: Once the training is complete and a fully-updated Q-table is available, the action with the highest Q-value from the present state is chosen.

\subsection{L-step look-ahead Q-Learning}
We noted from the trend study that there is a spike in error a few transmission before the loss of trajectory. To reinforce this, we propose a modification of Q-learning called the $L$-step lookahead Q-learning where we explicitly update the Q-values of the $M$ previous states for each transmission. The reward, action and state formulations are same as vanilla Q-Learning and so is the action selection method. The $L$ update equations are represented by
\begin{align}
    Q[s_{t-m},a_{t-m}]&=Q[s_{t-m},a_{t-m}]+ \nonumber\\
    &\hspace{-0.5cm} \alpha(r_t + \gamma \max(Q[s_t,:]) - Q[s_{t-m},a_{t-m}]),
\end{align}
for $m = 1,\cdots, L$. Here, $\alpha$ and $\gamma$ denote the learning rate and discount factor respectively.

\section{Simulation Results}
To demonstrate the performance of our proposed methods, we present results on a synthetic trajectory. \revi{We use the tool \textit{RocketPy} to generate the synthetic trajectory\footnote{The parameters of the synthetic  trajectory are chosen based on real-life data, which we are not at liberty to share.}.} We also include performance of fixed bandwidths of 1 and 5 MHz to compare with our proposed adaptive selection methods.
\begin{figure}[ht]
    \centering
    \includegraphics[width = 0.7\columnwidth]{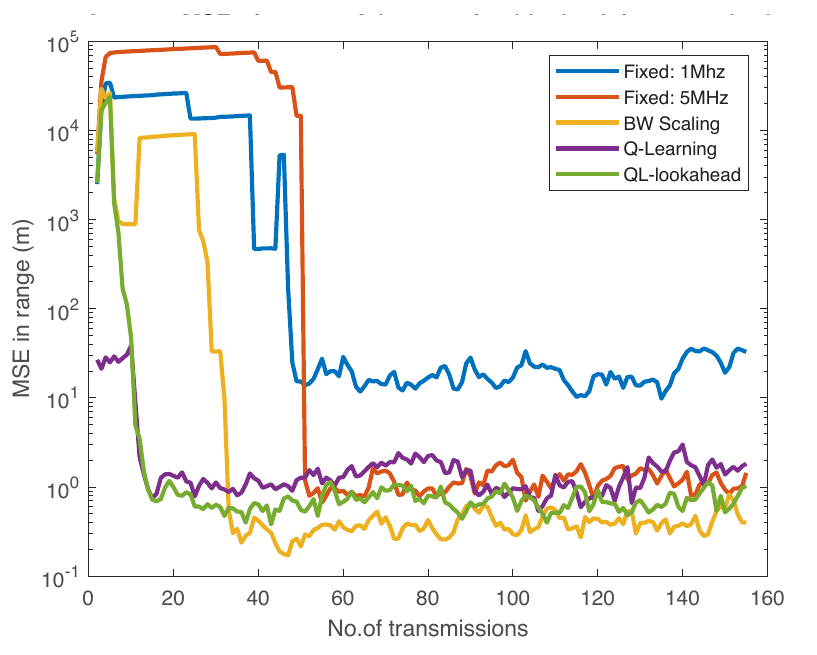}
    \caption{Average MSE in range over 100 runs vs. No. of transmitted beams illuminated on target}
    \label{fig:mse_blockmin}
\end{figure}

\begin{figure*}
\centering

\subfloat[Fixed BW: 1MHz]{\label{fig:win_bw1} \includegraphics[width=0.28\textwidth]{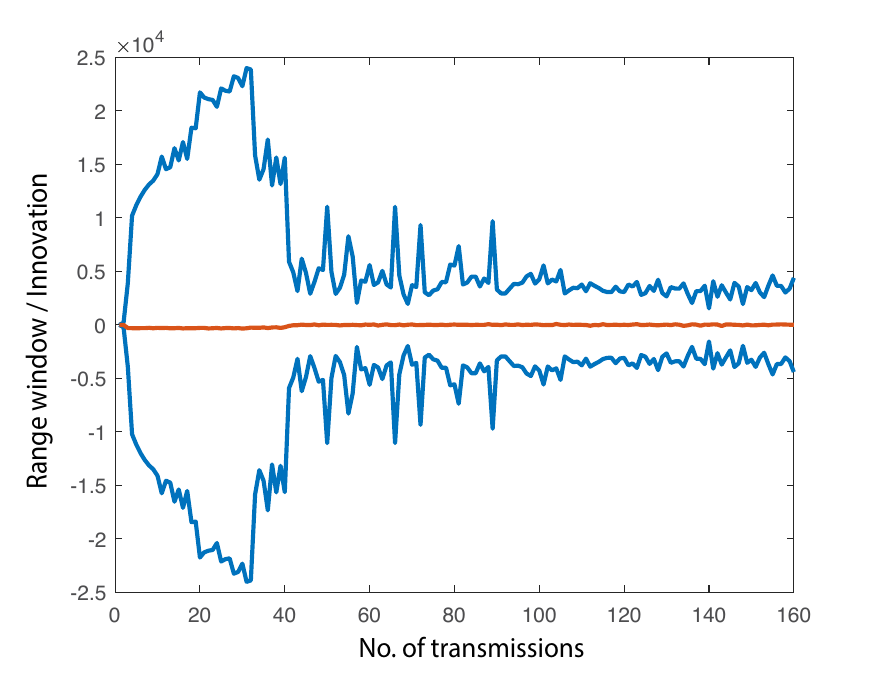}}%
\hfill
\subfloat[Bandwidth scaling]{\label{fig:win_bwsc} \includegraphics[width=0.28\textwidth]{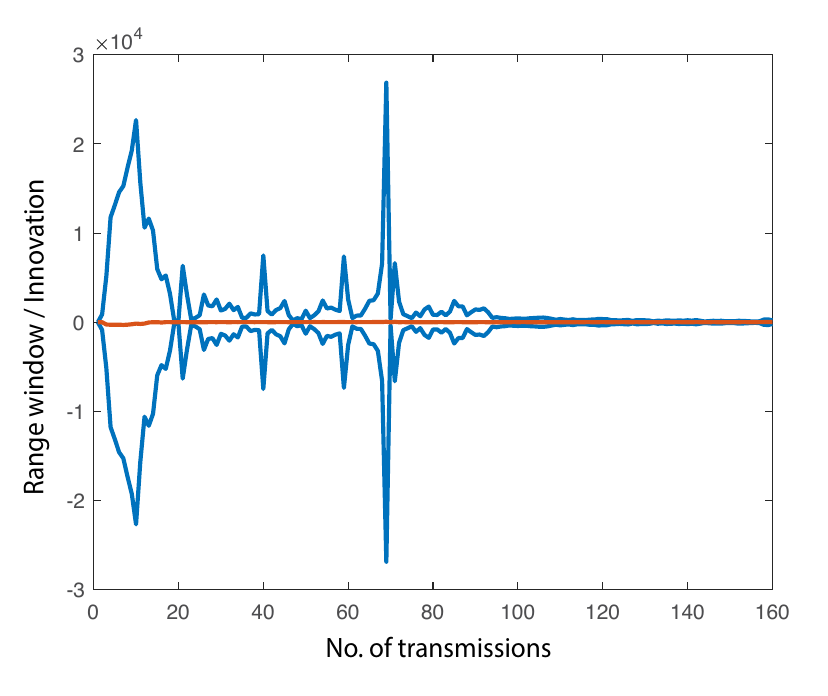}}%
\hfill
\subfloat[Q-Learning lookahead]{\label{fig:win_ql} \includegraphics[width=0.28\textwidth]{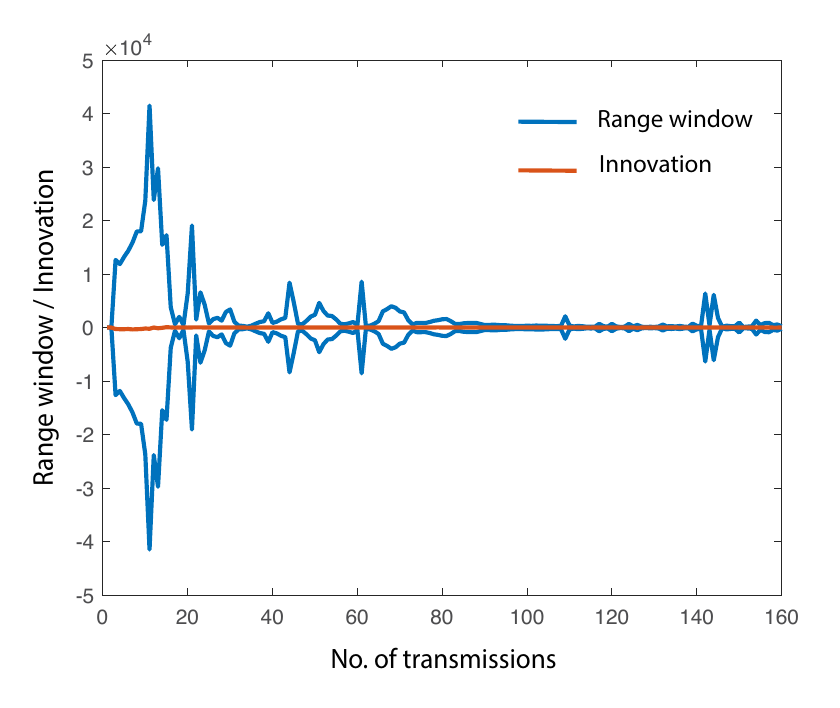}}%
        \caption{Range window and innovation vs. No. of transmissions}
        \label{fig:three graphs}
\end{figure*}

 \subsection{Parameters of Q-Learning}
 For this implementation of Q-learning, we consider 10 intervals of prediction error variance and 8 intervals of measurement error variances. Therefore, the state space is of dimension 80. We also consider the following action set
 \begin{equation}
     \mathcal{A} = \{0.5, 1, 2.5 , 5, 7.5,10\} \text{MHz}.
 \end{equation}
 Therefore, the Q-table consists of 480 entries to be learnt. A learning rate of $0.1$ and discount factor of $0.9$ is employed.
 For formulating the reward, the constant $C$ in (\ref{eqn:rewardql}) is chosen to be 2. The Q-table is updated for 200 runs, each consisting of 160 transmissions. For the lookahead method, we use $L=5$.
\begin{figure}[!ht]
    \centering
    \includegraphics[width = 0.7\columnwidth]{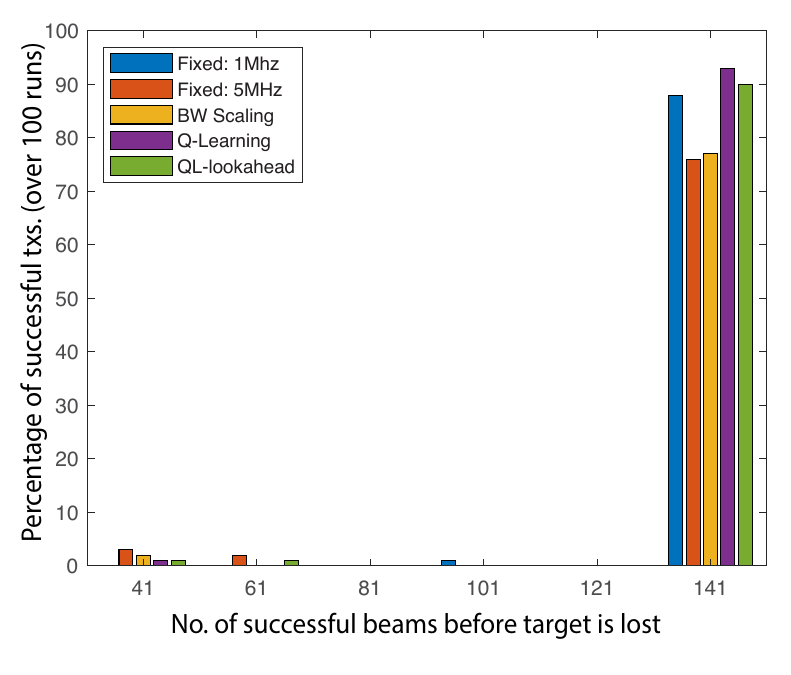}
    \caption{Histogram of the number of beams before the target was lost over 100 runs}
    \label{fig:hist}
\end{figure}

\subsection{Results for average behaviour}
The target was tracked across 160 instances, with the beam consistently placed on the target each time, and 100 such independent runs are conducted. Average Mean Square Error (MSE) in range over 100 runs  for successful transmissions is plotted in Fig. \ref{fig:mse_blockmin}. As MSE contains multiple peaks in a trajectory, a simple average over 100 runs does not provide any information. \rev{This is because the peaks may occur at different times in each of the runs; therefore, an average across runs does not provide meaningful insight into the trend of errors across the course of the transmission. Therefore, we consider the following metric: we choose a window of 3 consecutive transmission and consider the minimum of these values as a metric for our plots.} This is done to observe the performance while ignoring the temporary spikes. Histogram of successful transmissions is plotted in Fig. \ref{fig:hist}, which indicates at what stage of the trajectory was the target lost. (Note that the final bin in the figure corresponds to an entirely successful track.) From Fig. \ref{fig:mse_blockmin}, we note that the average MSE of successful transmissions is the lowest for BW scaling, however, it also has the lowest number of successful transmissions, as can be seen from Fig. \ref{fig:hist}. We note that constantly using a high bandwidth of 5MHz leads to loss of trajectory, as suggested by our initial trend study. Both Q-learning methods perform well in terms of MSE as well as the number of successful tracks, and provide a good trade-off between the two.

\subsection{Analysis for a single run}
To demonstrate establish the efficacy of the proposed approach, we plot the difference between measured and predicted range (referred to as innovation) \revi{as well as the acceptable range window against the number of transmissions for one run when the detection was successful throughout.} \revi{By range window, we mean the window of innovation that allows for detection.} For successful detection, the innovation value should not exceed 3 times the range window on either side (\revi{$95\%$ confidence interval computed from the measurement covariance}). An instance from a single trajectory is plotted for fixed bandwidths of 1MHz, Q-Learning lookahead, and Bandwidth scaling in Fig. \ref{fig:three graphs}. 
Using a low constant bandwidth of 1 MHz results in a wide window. The ideal situation of having a narrow window with the range innovation lying within is achieved using the proposed Q-Learning lookahead and bandwidth scaling. \rev{Note that although both methods perform similarly in this single instance analysis, Q-learning is a better method due to higher successful transmissions as indicated in Fig. \ref{fig:hist}.}

\subsection{Extension to other trajectories}
\rev{The same set of proposed methods were applied to an unseen trajectory. The Q-table trained for the previous trajectory is employed here as well, without any re-tuning. The average MSE for successful transmissions is plotted in Fig. \ref{fig:mse_blockmint2}. 
 \begin{figure}[h]
    \centering
    \includegraphics[width = 0.7 \columnwidth]{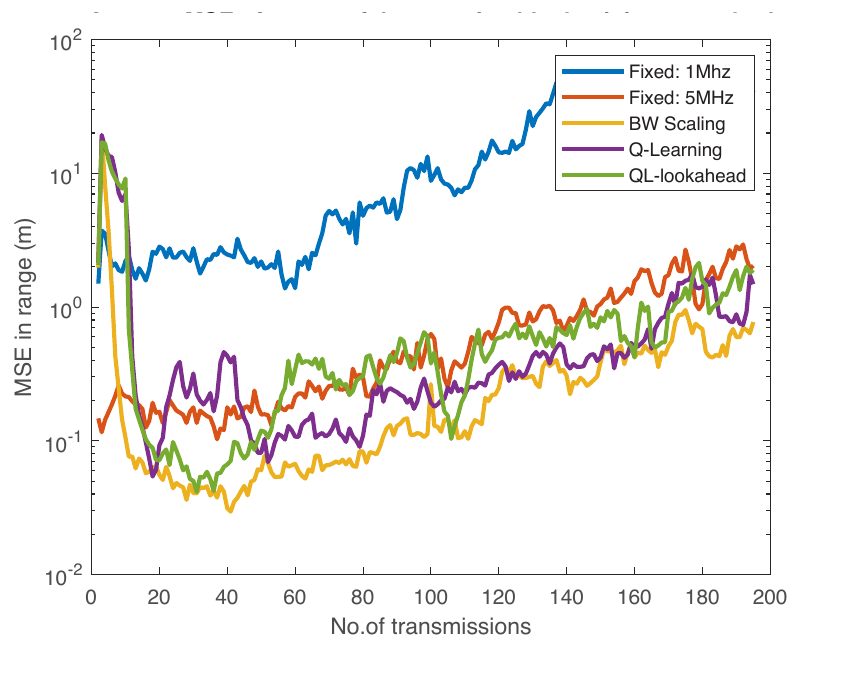}
    \caption{Average MSE in range over 100 runs vs. No. of transmitted beams illuminated on target for an unseen trajectory
    }
    \label{fig:mse_blockmint2}
\end{figure}
It can be seen that the proposed adaptive waveform selection methods of bandwidth scaling and Q-learning variants produce a lower error when compared to fixed bandwidths. This result shows that the proposed methods can be extended to other unseen trajectories without any retraining. As this trajectory is simpler to maneuver, all the methods yield successful transmissions over all runs. Therefore, the histogram is not plotted.}

\section{Conclusions}
In this work, we analysed the choice of adaptive waveform parameters to enhance the performance of cognitive radar. During analysis, we determined that the choice of bandwidth had more impact on the performance than the PRF. We proposed a few techniques for adaptive bandwidth selection: bandwidth scaling and Q-Learning (with and without lookahead). We noted that the Q-Learning with lookahead is able to outperform the existing and other proposed methods and can be extended to other trajectories without requiring any fine tuning.

	\bibliographystyle{IEEEtran}
    \bibliography{main.bib}

\end{document}